\documentclass[a4, 10pt]{article}

\usepackage{comment}
\usepackage{todonotes}

\usepackage{subcaption} 
\usepackage{caption}    
\usepackage{sidecap}
\usepackage[utf8]{inputenc}

\DeclareCaptionFormat{myformat}{\fontsize{9}{9}\selectfont#1#2#3}
\sidecaptionvpos{figure}{c}
\usepackage{cite}
\usepackage{textcomp} 

\usepackage{graphicx}

\usepackage[cmex10]{amsmath}
\usepackage{amssymb}

\usepackage{url}

\usepackage{listings}

\usepackage{caption}
\usepackage{subcaption}
\usepackage{mathtools}

\usepackage[]{units}

\usepackage{multirow} 
\usepackage{lipsum}
\usepackage{xcolor,colortbl}
\usepackage{tikz, tikz-3dplot}
\usetikzlibrary{positioning, patterns}
\usetikzlibrary{arrows,automata, shapes, petri}
\usetikzlibrary{arrows.meta, shapes.callouts}
\usetikzlibrary{calc,decorations.pathreplacing}
\usepackage{tikz-timing}

\usepackage{tabularx}

\usepackage{pgfplots}
\usepackage{booktabs}
\usepackage{pdfpages}

\usepackage[ampersand]{easylist}

\usepackage{amsthm}
\usepackage{amssymb}
\usepackage{dsfont}
\usepackage{bbm}

\newtheoremstyle{mystyle2}
{}
{}
{}
{}
{\itshape}
{.}
{ }
{}
\theoremstyle{mystyle2}

\theoremstyle{remark}

\usepackage{authblk}

\usetikzlibrary{decorations.pathreplacing}
\newcommand{\R}{\mathbb{R}}

\definecolor{col1}{RGB}{27,158,119}
\definecolor{col2}{RGB}{217,95,2}
\definecolor{col3}{RGB}{117,112,179}

\definecolor{MyLightGreen}{HTML}{66CC33}
\definecolor{MyPurple}{HTML}{9900CC}
\definecolor{MyLightBlue}{HTML}{0066FF}
\definecolor{tablegray}{rgb}{0.95,.95,.95}

\newcommand{\ie}{i.\,e.\ }

\newcolumntype{L}[1]{>{\raggedright\arraybackslash}p{#1}} 
\newcolumntype{C}[1]{>{\centering\arraybackslash}p{#1}} 
\newcolumntype{R}[1]{>{\raggedleft\arraybackslash}p{#1}} 
\newcolumntype{Z}{>{\raggedright\let\newline\\\arraybackslash}X}

\begin{document}

\title{Area Optimization with Non-linear Models in Core Mapping for System-on-Chips}

\author[1]{Jan~Moritz~Joseph}
\author[1]{Dominik Ermel}
\author[1]{Tobias Drewes}
\author[2]{Lennart Bamberg}
\author[2]{Alberto Garc\'ia-Oritz}
\author[1]{Thilo Pionteck}
\affil[1]{Otto-von-Guericke-Universit\"at Magdeburg\\
		Institut f\"ur Informations- und Kommunikationstechnik, 39106 Magdeburg, Germany\\
		Email: \{jan.joseph, dominik.ermel, tobias.drewes, thilo.pionteck\}@ovgu.de}
\affil[2]{University of Bremen\\
		Institute of Electrodynamics and Microelectronics, 28359 Bremen, Germany\\
		Email: \{agarcia, bamberg\}@item.uni-bremen.de}

\maketitle


\begin{abstract}
Linear models are regularly used for mapping cores to tiles in a chip. System-on-Chip (SoC) design requires integration of functional units with varying sizes, but conventional models only account for identical-sized cores. Linear models cannot calculate the varying areas of cores in SoCs directly and must rely on approximations. We propose using non-linear models: Semi-definite programming (SDP) allows easy model definitions and achieves approximately 20\% reduced area and up to 80\% reduced white space. As computational time is similar to linear models, they can be applied, practically. 
\end{abstract}


\section{Introduction}

A common design problem for System-on-Chips (SoCs) using mesh-based Network-on-Chips (NoCs) is core mapping. This takes a core graph and a network graph as input. In the core graph, nodes represent cores, edges represent communication and edge weights represent required bandwidths. In the network graph, nodes represent tiles, each of which is a NoC router with reserved space for a core, and edges represent links between tiles. The reserved area per core is usually identical. The output of core mapping is an assignment of cores to tiles. The objective function of core mapping minimizes communication costs, e.g.\ required bandwidth times hop distance. This is solved using linear models, for instance mixed-integer linear programming (MILP), or heuristics such as simulated annealing (SA), e.g. \cite{Srinivasan.2006}. 

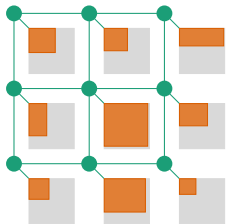
\begin{figure}[bth]
	\centering
	\begin{tikzpicture}

\foreach \x in {1,2}{
	\foreach \y in {1,2, 3}{
		\draw[col1] (\x, \y) -- (\x+1, \y);
	}
}
\foreach \x in {1,2, 3}{
	\foreach \y in {1,2}{
		\draw[col1] (\x, \y) -- (\x, \y+1);
	}
}
\foreach \x in {1, 2, 3}{
	\foreach \y in {1, 2, 3}{
		\draw[col1] (\x, \y) -- (\x+.2, \y-.2);
		\fill[col1] (\x, \y) circle (3pt);
	
	}
}	
\foreach \x in {1, 2, 3}{
	\foreach \y in {1, 2, 3}{
		\draw[fill = black!15, black!15] (\x+.2, \y-.2) rectangle ++ (.6, -.6);
}}

\foreach \x in {1, 2, 3}{
	\foreach \y in {1, 2, 3}{
	\draw[col2, fill=col2!80] (\x+.2, \y-.2) rectangle ++ (.2*rand+.4, -.2*rand-.4);
}}
\end{tikzpicture}
	\caption{Mapping of cores (orange) of varying size to a SoC with equal-sized tiles. Whitespace is gray.}
	\label{fig:mapping}
\end{figure}

The aforementioned approach does not account for cores with varying sizes due to different functionality, which is a typical characteristic of SoCs. Efficient core mapping for SoCs must account for this heterogeneity. An example for this inefficiency is depicted in Fig.~\ref{fig:mapping}, in which cores of different sizes (orange) allocate less area than reserved (light gray). Naturally, conventional linear models for mapping will be limited if the objective function also takes area into consideration, since area calculations are non-linear, intrinsically. 

In this publication, we propose to use non-linear models for area minimization during mapping, since these yield better area with short run time. We demonstrate this by comparison of a linear and a non-linear model for area optimization during core mapping: The proposed non-linear model yields better area, since an error-inducing linearization of area is not necessary. Further, they are cleaner than linear models and their definition is less complicated than usually anticipated. Plus, the additional computational effort for optimization is very small. Due to this, non-linear models are highly relevant in practical application.

This publication is structured as follows: First, we contribute two models for minimization of area for mapping in Sec.~\ref{sec:models}; one linear and one non-linear model. Next, we compare the results of linear and non-linear models in Sec.~\ref{sec:res:lp}. Then, the results of a mapping algorithm using SA, which considers area using non-linear models, are shown in Sec.~\ref{sec:res:rel} and compared to related work. Results are discussed in Sec.~\ref{sec:dis}. Finally, we conclude in Sec.~\ref{sec:con}.
\section{Area for mapping on cores}\label{sec:models}

We introduce two basic models, which allow optimizing the area of a SoC during core mapping. The first model is linear and therefore approximates area. The second is non-linear and therefore more area-efficient. These two models are the basis, on which we compare linear and non-linear models.

The area optimization problem is formulated as follows: Assume a given order of $l k$ or fewer components in $l$ rows and $k$ columns. This represents a given mapping of cores to tiles. Each component has the size $F_{i,j}$, for certain $i \in [l] := \{1, \dots l\} $ and $j \in [k] := \{1, \dots k\}$. $F_{i,j}=0$ if there is no component at row $i$, column $j$ for all pairs $(i,j) \in [l]\times[k]$. The height of rows is denoted by $r_i \in \R$ for all $i \in [l]$. The width of columns is denoted by $c_j \in \R$ for all $j \in [k]$. It holds that the reserved area in a tile is bigger than the size of the assigned core: 
\begin{equation}
r_i c_j \geq \text{F}_{i,j} \qquad \text{for all }i \in [l], j \in [k]. \label{eq:areanonlinear}
\end{equation}
The linearized objective function throughout this paper is to minimize the side length of a square that encloses all tiles.
\begin{equation}
\max\left(\sum_{i\in[l]} r_i, \sum_{j\in [k]} c_j\right) \longrightarrow \text{min}
\end{equation}

\subsection{Linear Model}

\begin{figure}[bth]
	\centering
	\begin{minipage}[b]{.670\linewidth}
		\includegraphics[width=\linewidth]{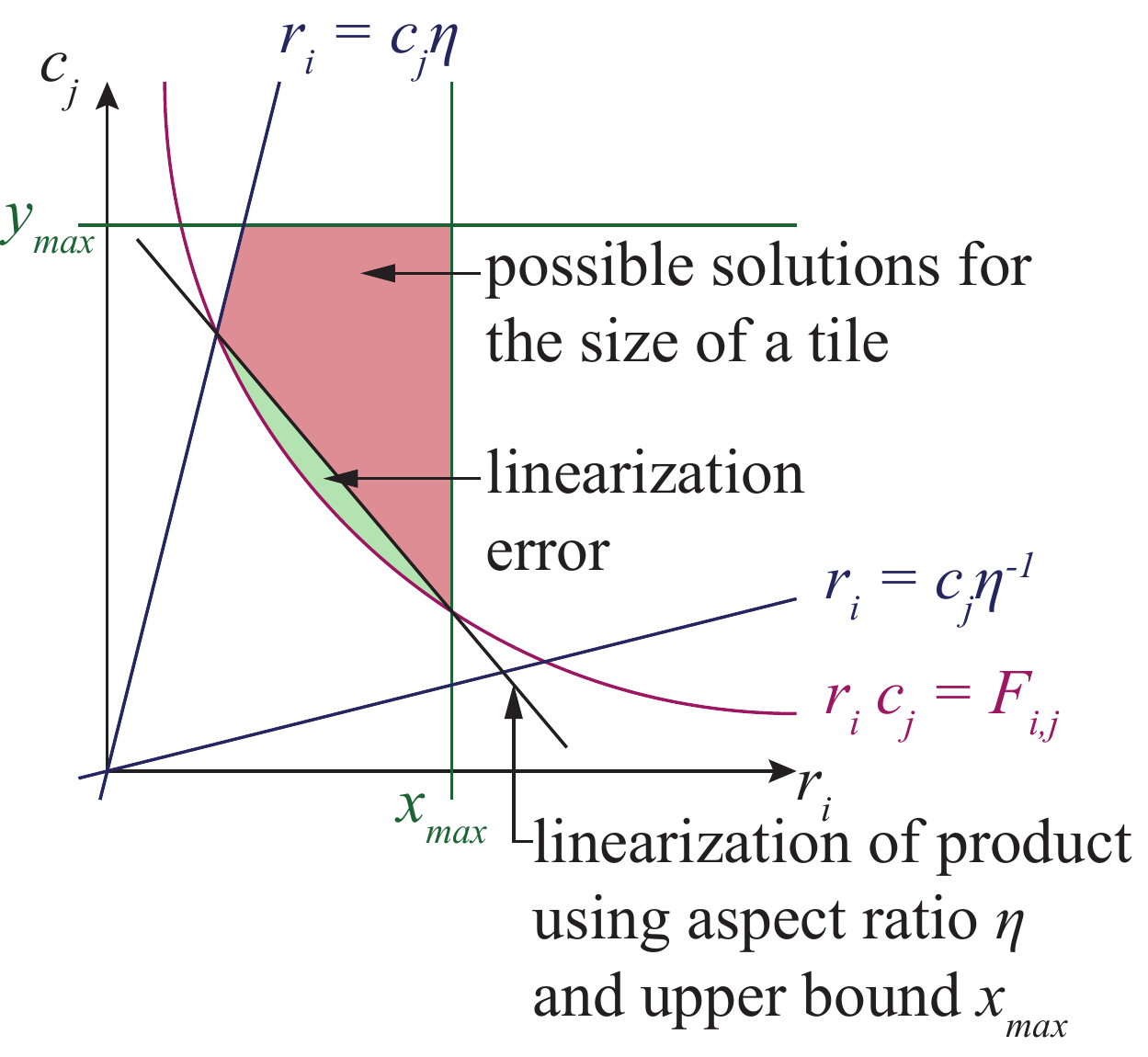}
		\subcaption{Simple approximation with single linear equation.}
		\label{fig:lin1}
	\end{minipage}\\
	\begin{minipage}[b]{.763\linewidth}
		\includegraphics[width=\linewidth]{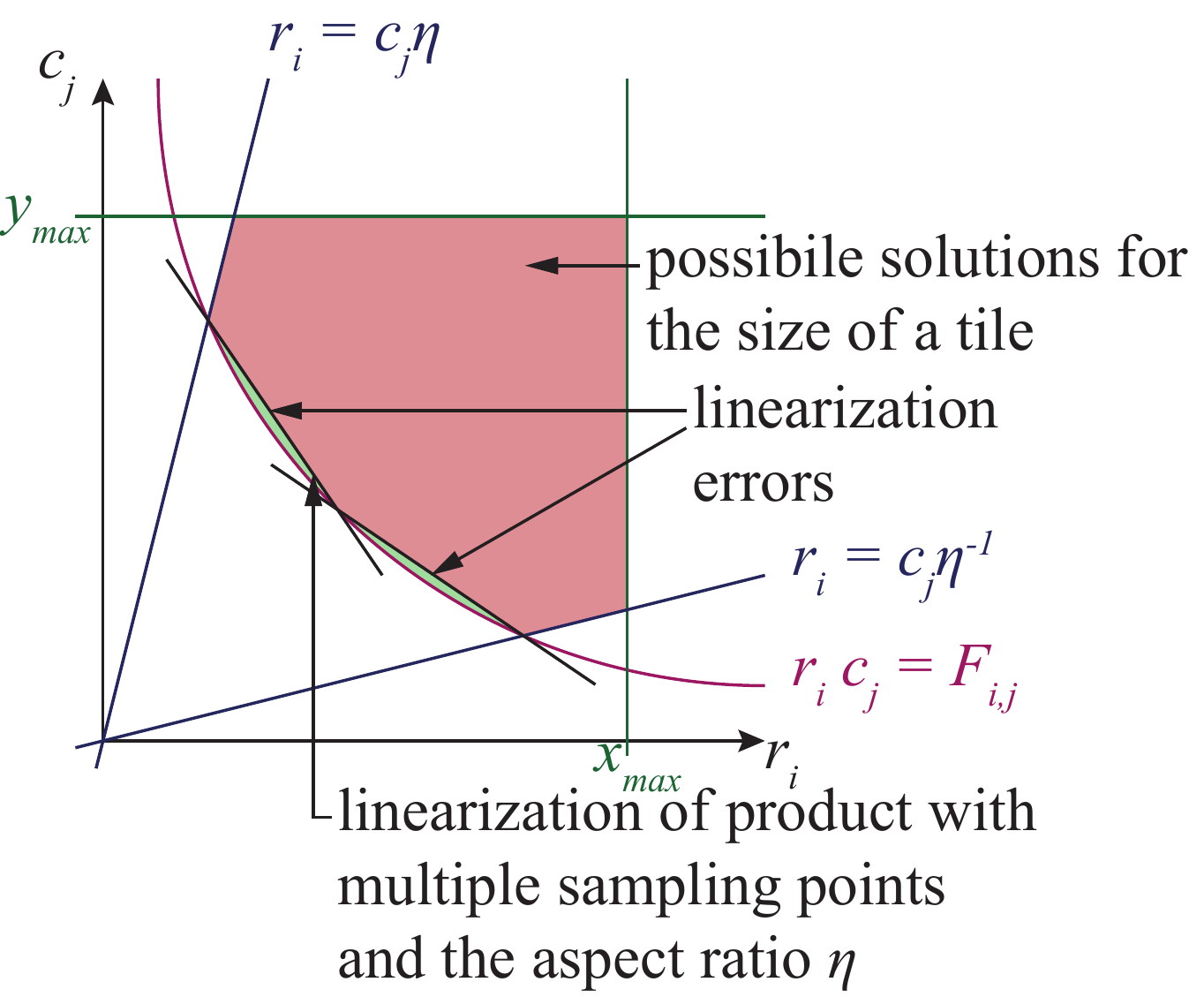}
		\subcaption{Reduced error through multiple linear approximations.}
		\label{fig:lin2}
	\end{minipage}
	\caption{Area linearization.}
	\label{fig:lin}
\end{figure}

The area of a rectangle $F_{i,j}$ with edge length $r_i$ and $c_j$ cannot be calculated through means of a linear model. Therefore, the area is linearized. A natural approach is given by \emph{Lacksonen et al.} \cite{Lacksonen.1994} for a factory layout problem, which can be applied here as well. The approach is shown in Fig.~\ref{fig:lin}. Linearization is possible, since rectangles are within an aspect ratio of $\eta \in (0,1)$, \ie $r_i \geq c_j \eta$ and $r_i \leq c_j \eta^{-1}$. This is shown in Fig.~\ref{fig:lin1}. The area $r_i c_i$ of a tile $i,j$ must be larger than its core with size $F_{i,j}$, \ie $F_{i,j} \leq r_i c_i$. The hyperbola for equal area is the lower left bound for the solution space of the optimization. Further, the solution space is limited by any given maximum edge length of the tile, \ie $c_j \leq y_{max}$ and $r_i \leq x_{max}$. The solution space is further reduced in size by the line equations for the aspect ratio $\eta$. Since the hyperbola is non-linear, it is approximated by a line equation given by the intersections between the lines for the aspect ratios and the maximum edge length. The resulting linearization error is plotted in green in Fig.~\ref{fig:lin1}. It is possible to reduce this error by including more equally-spaced knots as shown in Fig.~\ref{fig:lin2}. Each linear equation connecting two adjacent knots intersected with the iso-area-hyperbola ($r_ic_j=F_{i,j}$) is called a 1-spline. Please note that more 1-splines reduce the error but increase the model complexity, since integer inequalities are required to determine the current spline. There are at least three additional integer inequalities per supporting point. Naturally, this has a large performance impact.


For a given mapping, the optimization is subject to 
\begin{alignat}{2}
r_i & \geq \eta c_j \qquad &\forall i \in [l], \forall j \in  [k]\\ 
c_j & \geq \eta r_i \qquad &\forall i \in [l], \forall j \in  [k]\\ \label{eq:line}
r_i + c_j & \geq \sqrt{F_{i,j} \eta}+\sqrt{F_{i,j} / \eta} \qquad &\forall i \in [l], \forall j \in  [k]
\end{alignat}
with the minimum tile aspect ratio $\eta \in (0,1)$. The linearization of Eq.~\ref{eq:areanonlinear} is now Eq.~\ref{eq:line}, following the approach presented using only a single linear approximation (cf.~Fig.~\ref{fig:lin1}).

\subsection{Non-Linear Model}

Since linear models introduce a significant error, we use a semi-definite model to optimize the problem without this linearization error, because the red hyperbola in Fig.~\ref{fig:lin} is modeled. We set $k\, l$ variables $X_{k(i-1)+j}$ such that
\begin{equation}
X_{k(i-1)+j}=\left[\begin{array}{cc}
r_{i} & \sqrt{F_{i,j}}\\
\sqrt{F_{i,j}} & c_{j}
\end{array}\right]\succeq0,\ \forall i\in[l],\ \forall j\in[k]
\end{equation}
These matrices are premised to be positive semidefinite (\ie "$\succeq0$"); thus each principal minor is greater or equal to $0$:
\begin{align}
det&\left(X_{k(i-1)+j}\right)\geq 0 \\
\Leftrightarrow \quad r_{i}c_{j}&-F_{i,j}\geq 0\\
\Leftrightarrow \quad r_{i}c_{j}&\geq F_{i,j}, \qquad \forall i\in[l],\ \forall j\in[k]
\end{align}
We formulate a semi-definite programming (SDP) problem. The objective function minimizes a variable $x \geq \max \{\sum r_i, \sum c_i\}$ subject to:

We assign the corresponding area values to each matrix using the Frobenius inner product:
\begin{equation}
\begin{split}
2\sqrt{F_{ij}}  \leq & \left\langle \left[\begin{array}{cc}0 & 1\\1 & 0\end{array}\right],\ X_{k(i-1)+j}\right\rangle \\ \leq & \enspace 2\sqrt{F_{ij}}, \qquad \forall i \in [l], \forall j\in[k]
\end{split}
\end{equation}
For each $i \in [l]$, the upper left entry of the matrices $X_{k(i-1)+j}$ has the same value for all $j\in[k]$ (this models $r_i$): 
\begin{equation}
\begin{split}
0 \leq &  \left\langle \left[\begin{array}{cc}1 & 0\\0 & 0\end{array}\right],\ X_{k(i-1)+1}\right\rangle +  \\ &  \left\langle \left[\begin{array}{cc}-1 & 0\\0 & 0\end{array}\right],\ X_{k(i-1)+j}\right\rangle  \leq  0
\end{split}
\end{equation}
For each $j \in [k]$, the lower right entry of the matrices $X_{k(i-1)+j}$ has the same value for all $i\in[l]$ (this models $c_j$): 
\begin{equation}
\begin{split}
0  \leq & \left\langle \left[\begin{array}{cc}0 & 0\\0 & 1\end{array}\right],\ X_{j}\right\rangle + \\ & \left\langle \left[\begin{array}{cc}0 & 0\\0 & -1\end{array}\right],\ X_{k(i-1)+j}\right\rangle   \leq0
\end{split}
\end{equation}
We model the maximum variable $x$ for the objective function (this models $x \geq \sum r_i$ and $x\geq \sum c_i$):
\begin{align}
0 & \leq x+\sum_{i=1}^{l}\left\langle \left[\begin{array}{cc}-1 & 0\\0 & 0\end{array}\right],\ X_{k(i-1)+1}\right\rangle \\
0 & \leq x+\sum_{j=1}^{k}\left\langle \left[\begin{array}{cc}0 & 0\\0 & -1\end{array}\right],\ X_{j}\right\rangle  
\end{align}
Again, areas of tiles are constrained by an aspect ratio $\eta$. Note, that this aspect ratio is not violated by the relation between $r_i$ and $c_j$. Rather, a component can find a rectangle inside the bounding box given by $r_i c_j$. This rectangle has the size of the core. The aspect ratio of its edges is greater than $\eta$. We formulate for all $i \in[l]$ and for all $j \in [k]$:
\begin{align}
\sqrt{\eta F_{i,j}} &\leq \left\langle \left[\begin{array}{cc}1 & 0\\0 & 0\end{array}\right],\ X_{k(i-1)+1}\right\rangle \\
\sqrt{\eta F_{i,j}} &\leq \left\langle \left[\begin{array}{cc}0 & 0\\0 & 1\end{array}\right],\ X_{j}\right\rangle
\end{align}

\vspace{3pt}
\section{Results}

We first compare the provided linear and non-linear model using theoretical benchmarks to quantify advantages for area and analyze additional effort in computational time in Sec.~\ref{sec:res:lp}. Then, we apply the non-linear model to mapping of cores with different sizes in a SoC and compare the results to the related work in Sec.~\ref{sec:res:rel}. 

\subsection{Linear vs.\ non-linear model}\label{sec:res:lp}

We compare the results of linear and non-linear models using the proposed LP and SDP. The models are implemented in MATLAB R2018a. LPs are optimized using IBM CPLEX 12.8.0 \cite{Cplex.2018}. SDPs are optimized using Mosek 8.1 \cite{MosekApS.2018}.

We generate three input benchmarks. The cores are equal-sized to provide a fair comparison against conventional approaches. The benchmarks are: 
\begin{enumerate}
	\item A 3D SoC with 2 layers and 5 tiles, of which three tiles are in layer 1 and two tiles are in layer 2.
	\item A 3D SoC with 4 layers and 10 tiles per layer connected by a 2$\times$5 mesh NoC. 
	\item A 3D SoC with 4 layers and 20 tiles per layer connected by a 4$\times$5 mesh NoC.
\end{enumerate}
Cores are \unit[10]{A} large. Routers with 5 ports require \unit[1]{A}. The router area is linearly proportional to port count depending on the position of the router in the network. TSV arrays, which vertically connect routers, are \unit[2]{A} large. The aspect ratio is limited by $\eta = 0.1$. We run the optimization 50 times to average run time on an Intel Core i7-6700 (4 physical cores at \unit[3.4]{GHz} with hyper threading) using Windows 10.

\begin{table}
		\caption[Area LP vs.\ SPD]{Area, runtime, inequality count and variable count comparison between linear and non-linear model to optimize a homogeneous 3D SoC with 5, 40 and 80 homogeneous cores including TSV areas (runtime average of 50 reruns).}
	\label{tab:area}
	\setlength\tabcolsep{5.5pt} 
	{\scriptsize
		\begin{tabularx}{1\linewidth}{p{.05cm}|r|r|r|r|r|r|r|r|r} 
			\toprule
			\parbox[t]{2mm}{\multirow{3}{*}{\rotatebox[origin=c]{90}{\textsc{Layer}}}} & \multicolumn{9}{c}{\textsc{Area [A] and Difference}} \\
			 & \multicolumn{3}{c|}{\textsc{5 PEs}}& \multicolumn{3}{c|}{\textsc{40 PEs}}& \multicolumn{3}{c}{\textsc{80 PEs}}\\
			&\textsc{LP} &\textsc{SDP} & \textsc{$\Delta$} &\textsc{LP} &\textsc{SDP} &\textsc{$\Delta$} &\textsc{LP} &\textsc{SDP} &\textsc{$\Delta$}\\
			\midrule
			\textsc{1} &43.0 & 36.8 &-14.4\%&211 & 178 &-15.6\%&364 & 301 &-17.4\%\\
			
			\textsc{2} &25.7 & 23.0 &-10.5\%&222 & 180 &-18.9\%&379 & 313 &-17.4\%\\
			
		 	\textsc{3} & --- & --- & --- &214 & 183 &-14.5\%&378 & 313 &-17.2\%\\
			
			\textsc{4} & --- & --- & --- &185 & 154 &-16.8\%&316 & 261 &-17.4\%\\
			
			\midrule
			
			\multicolumn{3}{l}{\textsc{Average Area}} & \multicolumn{1}{r|}{}& \multicolumn{3}{r|}{}& \multicolumn{3}{r}{}\\
			\multicolumn{3}{l}{\textsc{Reduction}} & \multicolumn{1}{r|}{-12.5\%}& \multicolumn{3}{r|}{-16.5\%}& \multicolumn{3}{r}{-17.3\%}\\
			
			\midrule
			
			 \multicolumn{1}{c}{}& \multicolumn{9}{c}{\textsc{Average runtime [s]}} \\
			\midrule
			 \multicolumn{2}{r|}{0.4} & 2.9 & +625\%& 3.9 & 7.5 & +92.3\%& 12.2 & 16.0&+31.1\%\\
			 
			 \midrule
			 			
			 \multicolumn{1}{c}{}& \multicolumn{9}{c}{\textsc{Inequality count}} \\
			 \midrule
			 \multicolumn{2}{r|}{16} & 31 & +94\%& 88 & 436  & +395\%& 168 & 1644& +879\%\\
			 
			 \midrule
			 
			  \multicolumn{1}{c}{}& \multicolumn{9}{c}{\textsc{Variable count and Difference [\%]}} \\
			 \midrule
			 \multicolumn{2}{r|}{9} & 21 & +133\%& 32 & 112 & +250\%&  40 & 200&+400\%\\
			
			
%
			\bottomrule
	\end{tabularx}}
\end{table}

\begin{table*}[t]
		\caption[Layer planning]{Area and network performance comparison of mapping to a 2D-mesh NoC with \cite{Srinivasan.2006}. The SA is executed with 20 reruns, an initial temperature of 30, cooling of 0.98 and 15,000 iterations. The aspect ratio is limited by $\eta = 0.1$.} 
	\label{tab:tp2-ref1}
	{\scriptsize
		\begin{center}
		\begin{tabularx}{.81\linewidth}{p{.1cm}p{.1cm}|Z||r|r|r||r|r|r||r|r|r} 
			\toprule
			 &&& \multicolumn{3}{c||}{\textsc{Area [A]}}  & \multicolumn{3}{c||}{\textsc{Communication [Bits*Hops]}} & \multicolumn{3}{c}{\textsc{Bandwidth [Bits]}} \\
			 &&& \textsc{mean}  & \textsc{std} & \textsc{ratio} & \textsc{mean}  & \textsc{std} & \textsc{ratio} &  \textsc{mean}  & \textsc{std}  & \textsc{ratio} \\
		
			\midrule
			\multirow{4}{*}{\rotatebox[origin=c]{90}{H256 \textsc{dec}}}& \multirow{4}{*}{\rotatebox[origin=c]{90}{mp3 \textsc{dec}}} & \textsc{Baseline \cite{Srinivasan.2006}} & 11301 & ---&---& 19858 &---&---& 4060 &---&---\\
			&&\textsc{Baseline with SDP}&	10178&---&	-9.94\%&19858&---	&0.0\%	&4060&---&	0.0\%\\
			&&\textsc{initial solution}	&7902&---&	-30.1\%&	33707&---&	+69.7\%&	7994&---&	+96.9\%\\
			&&\textsc{SA with SDP} &	8244&505&-27.1\%&21280&624&+7.16\%&4452&674&+9.66\%\\		
			\midrule
			
			\multirow{4}{*}{\rotatebox[origin=c]{90}{H263 \textsc{enc} }}&\multirow{4}{*}{\rotatebox[origin=c]{90}{mp3 \textsc{dec}}}&\textsc{Baseline \cite{Srinivasan.2006}} &12535 &---&---&255324&---&---&	84884&---&---	\\	
			&&\textsc{Baseline with SDP}&	10178&---&	-18.8\%&255324	&---&0.0\%&	84884&---&	0.0\%\\			
			&&\textsc{initial solution}	&6993&---&-44.2\%&525537&---&+106\%	&85244&---&+0.42\%\\	
			&&\textsc{SA with SDP} 	&10474&2148&-16.4\%&250187&14763&-2.0\%&73161&17497&-13.8\%\\

			\midrule
			
			\multirow{4}{*}{\rotatebox[origin=c]{90}{mp3 \textsc{enc}}}&\multirow{4}{*}{\rotatebox[origin=c]{90}{mp3 \textsc{dec}}}&\textsc{Baseline \cite{Srinivasan.2006}}&8568	&---&---&17546	&---&---& 4085&---&---\\
			&&\textsc{Baseline with SDP}&8091	&---&-5.57\%&17546&---&	0.0\%	&4085&---&	0.0\% \\
			&&\textsc{initial solution}&7281&---	&-15.0\%	&39171&---&	+123.3\%&	6560&---&	+60.1\%\\
			&&\textsc{SA with SDP}  &8516	&796& -0.61\%&17572&487&+0.15\%&4974&902&+21.8\%\\
			
			\bottomrule
	\end{tabularx}
	\end{center}
	}
\end{table*}

Both models require approximately the same amount of memory in MATLAB. The area and runtime results are reported in Table~\ref{tab:area}. In benchmark 1), the summed chip area is \unit[68.7]{A} from the LP and \unit[59.8]{A} from the SDP. In benchmark 2), the summed chip area is \unit[832]{A} from the LP and \unit[695]{A} from the SDP. In benchmark 3), the summed chip area is \unit[1272]{A} from the LP and \unit[1188]{A} from the SDP. Since in the lowest layer there is no TSV area required (there are no keep-out-zones using via-middle-process-flow), this layer is smaller. The difference in run time between LP and SDP is smaller for larger input sets.

The linear model requires $2kl + 2$ inequalities and $k+l+1$ variables. The non-linear model requires $(kl)^2 + k +l+2$ inequalities and $kl+1$ variables.

\subsection{Mapping with linear vs.\ non-linear models}\label{sec:res:rel}

Quadratic-shaped cores of a different size are mapped in a 2D mesh NoC in \cite{Srinivasan.2006} using a linear model and a heuristic for larger input sets. Their objective function does not target low area but minimizes transmission energy. We compare the results from our non-linear model with results from linear models from \cite{Srinivasan.2006} using the three benchmarks provided (see Table~\ref{tab:tp2-ref1}). The data streams for the benchmarks are taken from \cite{Sahu.2013} and the cores' area from \cite{Srinivasan.2006}. 

We implement a simple SA for mapping: The neighbor function randomly changes the position of one core in the mapping. If the position is already taken by another core, both will be switched. We extend the objective function from \cite{Srinivasan.2006} to include chip area. The initial solution for our SA places cores decreasingly ordered by size, such that white space is minimized and communication is not accounted for.

The results for area and network performance are given in Table~\ref{tab:tp2-ref1}. The sum of all tiles' sizes gives the total area (including whitespace). The network performance is measured by accumulated link load (measuring delay) and maximum link load (measuring throughput). Four data sets are given per benchmark: First, the baseline from \cite{Srinivasan.2006} for a 2D mesh NoC. Second, the first configuration is optimized using the proposed non-linear model without changing the mapping. This quantifies the potential of non-linearity. Third, the initial solution for the SA is given, which is area-efficient and communication-inefficient. Fourth, our SA is executed 20 times with 15,000 iterations, an initial temperature of 30 and a cooling of 0.98. The aspect ratio is limited by $\eta = 0.1$ The results of all runs are averaged and the standard deviation is calculated. We balance the weights in the cost function and prioritize neither area nor communication. A single run of the SA terminates after approximately 17 minutes on a Windows 10 workstation using an Intel i7-7740X processor (4 physical cores at \unit[4.3]{GHz} with hyper threading). 

\section{Discussion}\label{sec:dis}

\subsubsection{Area}

The area of non-linear models is better, as expected. The theoretical benchmarks, see Table~\ref{tab:area}, show their clear advantage in terms of area minimization. Area is reduced between 10.5\% and 18.9\%. The results for the real-world based benchmarks further support this because the area of the mapping is reduced by 5.57\% to 18.8\% in comparison to the solution from \cite{Srinivasan.2006}, see Table~\ref{tab:tp2-ref1}. If declined communication efficiency is acceptable, the area can further be reduced. For instance, the H256 dec mp3 dec can be mapped with 27.1\% area reductions at the expense of 7.16\% worse communication costs and 9.66\% worse bandwidth over baseline. The other two benchmarks show similar results. Summing up, non-linear models enable area reductions of up to 27\% in our benchmarks.
\subsubsection{Model definition}

The proposed model definition supports, that non-linear models are cleaner than linear models: First, error-prone linearization is not required. Second, the SDP directly models the area multiplication, which increases readability and allows for easier understanding of the problem. Therefore, the model is less complicated. Further, the effort to define a non-linear model is small. The whole model is given by 10 inequalities.
\subsubsection{Optimization run time}

Higher expected run time is a common reason to reject non-linear models. Our results do not support this. Of course, the run time of the non-linear model is slightly worse than the linear model. But the SDP only requires 16 seconds to find an optimal solution even for the large example with 80 cores. The variable count grows fast for the non-linear model (SDP: $\Omega((kl)^2)$ vs.\ LP: $\Omega(k+l)$) but this does not have a negative impact on performance. The number of inequalities grows similar for both at the rate of $\Omega(nm)$ and the optimizer for the SDP handles this better. The run time of the SA in the real-world based benchmarks of 17 minutes is reasonable for daily usage. This demonstrates that non-linear models are applicable in practice.

\section{Conclusion}\label{sec:con}

In this paper we show that using non-linear models for area optimization in core mapping yields better area with better-structured model definitions than conventional linear models. We contribute both a linear model (LP) and a non-linear model (SDP) to reduce the amount of white space in reserved areas for cores. Comparison of the two models shows that area can be reduced by up to 27\%. Since the non-linear model does not require area approximation, the model definition is cleaner. A small set of inequalities directly describes the problem completely. For a large benchmark example with 80 cores, an optimal solution is found within 16 seconds. 

\section*{Acknowledgments}
\noindent This work is funded by the German Research Foundation (DFG) project PI 447/8-1.


\bibliographystyle{IEEEtran}
\bibliography{bibliography}

\end{document}